\documentclass[twocolumn]{aastex701}

\usepackage{hyperref}

\begin{document}

\title{The extended globular cluster system of the archetypal ``failed galaxy'' Dragonfly-44 from deep white-light \textit{Hubble Space Telescope} imaging}

\author[orcid=0000-0003-3153-8543,gname=Bosque,sname='North America']{Maria Luísa Buzzo}
\affiliation{Astronomy Department, Yale University, 219 Prospect St, New Haven, CT 06511, USA}
\email[show]{marialuisa.buzzo@yale.edu}  

\author[orcid=0000-0002-8282-9888,gname=Bosque, sname='Sur America']{Pieter van Dokkum} 
\affiliation{Astronomy Department, Yale University, 219 Prospect St, New Haven, CT 06511, USA}
\affiliation{Dragonfly Focused Research Organization, 150 Washington Avenue, Suite 201, Santa Fe, NM 87501, USA}
\email{pieter.vandokkum@yale.edu}

\author[orcid=0000-0002-4542-921X,gname=Bosque, sname='Sur America']{Roberto Abraham} 
\affiliation{Department of Astronomy \& Astrophysics, University of Toronto, 50 Saint George Street, Toronto, ON M5S 3H4, Canada}
\affiliation{Dragonfly Focused Research Organization, 150 Washington Avenue, Suite 201, Santa Fe, NM 87501, USA}
\email{roberto.abraham@utoronto.ca}

\author[orcid=0000-0002-1841-2252,gname=Bosque, sname='Sur America']{Shany Danieli} 
\affiliation{School of Physics and Astronomy, Tel Aviv University, Tel Aviv 69978, Israel}
\email{sdanieli@tauex.tau.ac.il}

\author[orcid=0000-0003-2473-0369,gname=Bosque, sname='Sur America']{Aaron J Romanowsky} 
\affiliation{Department of Physics and Astronomy, One Washington Square, San José State University, San Jose, CA 95192, USA} 
\affiliation{Department of Astronomy and Astrophysics, University of California Santa Cruz, 1156 High Street, Santa Cruz, CA 95064, USA}
\email{aaron.romanowsky@sjsu.edu}

\begin{abstract}
For nearly a decade, Dragonfly-44 (DF44, $M_{\star} = 3\times10^8\,{\rm M}_\odot$) has been considered an archetypal ``failed galaxy'', a system so rich in globular clusters (GCs) and dark matter that it challenges standard dwarf galaxy formation scenarios. Yet a key measurement underpinning this classification has remained controversial, with published GC counts differing by a factor of four. Here we present new ultra-deep \textit{Hubble Space Telescope} WFC3/UVIS imaging of DF44 in the F350LP filter, reaching more than one magnitude below the canonical turnover of the GC luminosity function. We find that DF44 hosts $N_{\rm GC}=73.1\pm8.6$ GCs in a spatially extended system with a half-number radius of $R_{\rm gc}=1.48^{+0.39}_{-0.51}R_{\rm e}$, at the high end of previously published values. The GC system comprises ${\sim}3.9\%$ of the total stellar mass and, if DF44 follows the empirical GC number--halo mass relation, implies $\log(M_{\rm vir}/M_\odot)=11.5\pm0.3$, consistent within the uncertainties with the cored halo mass inferred from stellar kinematics by \cite{vanDokkum_19_DF44}. Together with a specific frequency of $S_N=42.1\pm4.9$, the rich GC population strongly supports the interpretation of DF44 as a failed-galaxy candidate that assembled a massive halo and rich GC population early, but never formed the field stellar mass expected for its halo. Current formation models reproduce several individual properties of DF44, but its combination of GC richness, early quenching, and high inferred halo mass remains an important constraint on models of dwarf-galaxy formation.
\end{abstract}


\section{Introduction}
\label{sec:intro}

Studies of ultra-diffuse galaxies in the past decade have consistently shown that the connections between galaxy size, stellar mass, dark matter content, and star-cluster formation are not as simple as we once thought \citep{Gannon_26}. First identified in large numbers in the Coma cluster \citep{vanDokkum_15}, UDGs are usually selected to have effective radii $R_{\rm e}>1.5\,{\rm kpc}$ and central surface brightnesses fainter than $\mu_{g,0}=24\,{\rm mag\,arcsec^{-2}}$. Their diffuse stellar bodies can look remarkably similar, yet they span an extraordinary range of physical properties. Some are gas rich and actively forming stars, while others are old, metal poor, and have been quiescent for most of their lives \citep{Roman_17,FerreMateu_23,Buzzo_24}. Some are strongly dark matter dominated within their stellar bodies, while others appear to contain very little or no dark matter over the same scales \citep{Beasley_16,vanDokkum_18,vanDokkum_19,Danieli_19,vanDokkum_19_DF44,Shen_23,Buzzo_25b,Buzzo_26}. Their globular cluster (GC) systems are similarly diverse: GCs are nearly absent in some UDGs, unusually numerous in others, and exceptionally luminous in several of the dark matter-deficient systems. UDGs have therefore become an important test of several aspects of dwarf-galaxy formation, rather than representing a single class of diffuse galaxies \citep{Gannon_26}.

Recent work has begun to connect this diversity to different formation pathways. UDGs that follow the dwarf-galaxy mass--metallicity relation tend to be younger, more elongated, GC poor, and found in lower-density environments, broadly consistent with ``puffed-up dwarf'' scenarios driven by high halo spin, stellar feedback, or environmental effects \citep{Amorisco_16,diCintio_17,Chan_18,Buzzo_22b,Buzzo_24,Buzzo_25a}. In contrast, GC-rich UDGs tend to be older, more metal poor, rounder, and quenched earlier, and are often interpreted as ``failed-galaxy'' candidates that formed efficiently at early times but stopped building their field-star populations soon afterward \citep{Peng_16,Danieli_22,FerreMateu_23,Buzzo_24,Forbes_25}.

GC systems are central to distinguishing between these pathways because they preserve several pieces of information about their hosts. The total number of clusters, $N_{\rm GC}$, and the specific frequency, $S_N$, measure whether or not a galaxy has an unusual number of GCs for its luminosity. The empirical $N_{\rm GC}$--halo mass relation provides an indirect estimate of halo mass \citep{Harris_17,Burkert_Forbes_20}, while $M_{\rm GC}/M_\star$ measures how much of the surviving stellar mass is contained in old bound clusters. The GCLF and spatial distribution provide further constraints on how the system formed and evolved \citep{Saifollahi_22,Forbes_25}. Yet, for some of the galaxies where these measurements matter most, even the total number of GCs has remained uncertain.

One such galaxy is DF44. It is among the largest UDGs known, with $R_{\rm e}=3.8\,{\rm kpc}$ \citep{Buzzo_25a}, and its old, metal-poor stellar population and short formation timescale suggest that most of its stellar mass formed early \citep{Villaume_22,Webb_22,FerreMateu_23}. Spatially resolved Keck/KCWI spectroscopy shows that DF44 is dispersion supported, has no detected rotation, and is strongly dark matter dominated within its stellar body \citep{vanDokkum_19_DF44}. Although the kinematics establish that the galaxy is dark matter dominated, its total halo mass depends on the adopted dark matter profile and stellar orbital anisotropy. The GC population therefore provides a valuable independent constraint. Yet the number of GCs needed for that estimate is precisely what has remained controversial.

\citet{vanDokkum_16} inferred approximately 100 GCs from Gemini imaging, and \citet{vanDokkum_17} obtained $N_{\rm GC}=74\pm18$ from \textit{HST}. However, \citet{Saifollahi_21} reanalysed the original \textit{HST} imaging and inferred a population of approximately 20 GCs. Using new \textit{HST} observations obtained through an independent program, \citet{Saifollahi_22} similarly found $N_{\rm GC}=20^{+6}_{-5}$ and a half-number radius of $R_{\rm gc}=0.78^{+0.44}_{-0.30}R_{\rm e}$, compared with the $R_{\rm gc}=1.5R_{\rm e}$ found by \citet{vanDokkum_17}. Despite the lower GC count, \citet{Saifollahi_22} found DF44 to lie at the high end of the dwarf-galaxy halo mass distribution and argued that its properties still favoured a failed massive-dwarf interpretation.

These are not small differences. Changing the inferred population from ${\sim}20$ to ${\sim}70$ GCs moves DF44 from the upper end of the dwarf-galaxy distribution to one of the most GC-rich systems known for its luminosity, with very different implications for $S_N$, $M_{\rm GC}/M_\star$, and halo mass. The goal now is to move beyond measuring the number and understand what it means for the formation of these galaxies. That is only possible once the number itself has been pinned down. At the distance of Coma, this requires imaging that reaches beyond the GCLF turnover, contains enough clusters to constrain their radial distribution, and samples enough of the surrounding field to characterize variations in the local background.

\begin{figure*}
\centering
\includegraphics[width=0.9\textwidth]{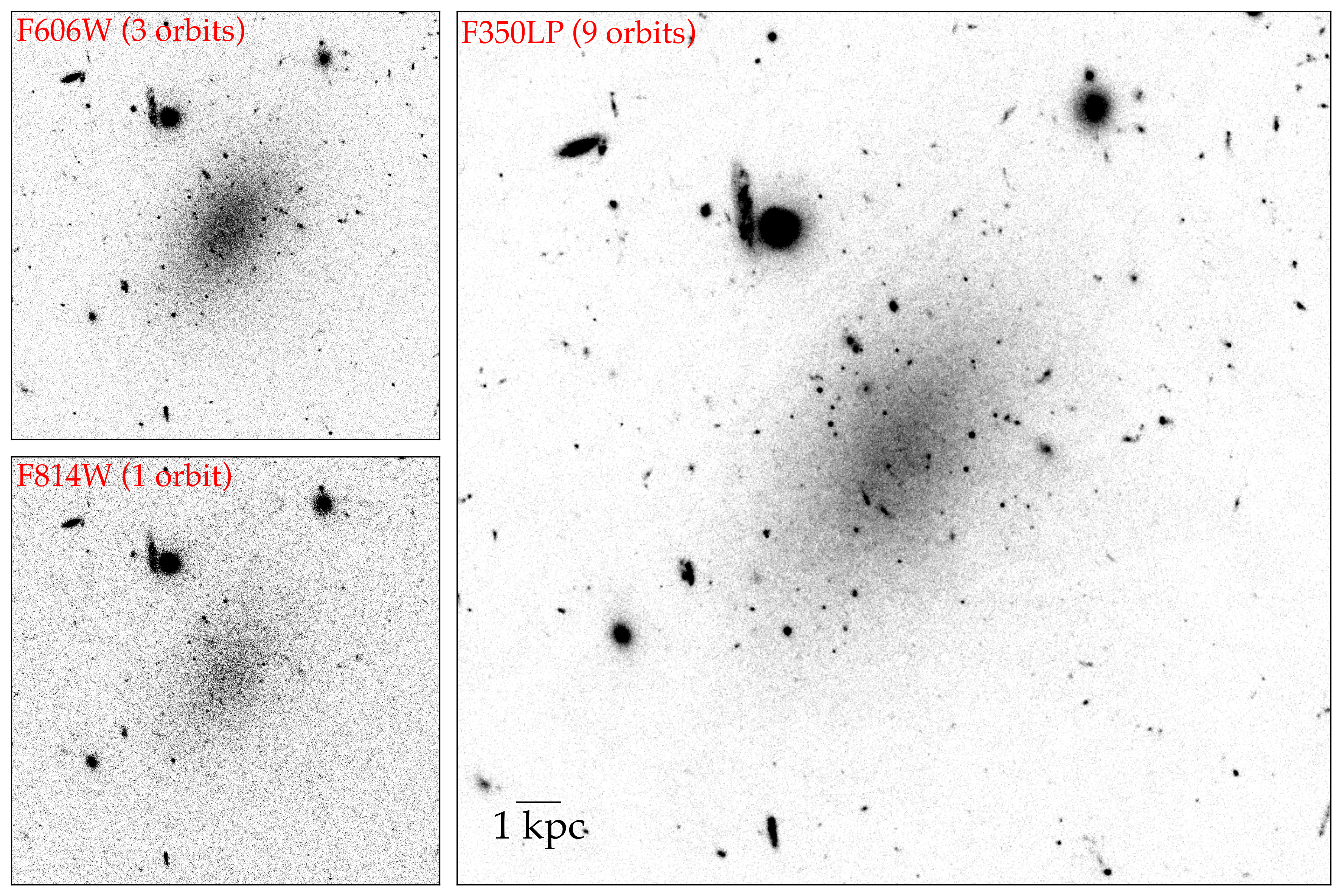}
\caption{Comparison between previous and new \textit{HST} imaging of Dragonfly~44. The left panels show the ACS imaging obtained by \citet{vanDokkum_17} and reanalysed by \citet{Saifollahi_21}: the upper panel corresponds to the combined F606W image and the lower panel to the shallower F814W image. The large panel on the right shows the new WFC3/UVIS F350LP data presented in this work. All panels are displayed using the same logarithmic stretch to facilitate a direct comparison of depth and surface brightness sensitivity. The new imaging is substantially deeper than the previous data and clearly detects a rich population of compact sources surrounding DF44. Each panel covers approximately $40''\times40''$, corresponding to $\sim20\times20\,{\rm kpc}$ at 100\,Mpc. North is up and East is to the left.}
\label{fig:DF44_HST}
\end{figure*}

In this paper, we present new ultra-deep \textit{Hubble Space Telescope} WFC3/UVIS imaging of DF44 in the broad F350LP filter. As shown in Figure~\ref{fig:DF44_HST}, the new data are substantially deeper than the previous imaging and reach more than one magnitude below the GCLF turnover. We use them to measure the background of compact sources, the radial distribution and luminosity function of the GC system, its total population, and its specific frequency. With these measurements in hand, we revisit what the GC system implies for the halo and formation history of DF44, moving the discussion from how many GCs the galaxy has to what its GC population means for the formation of GC-rich, dark matter-dominated dwarfs.

Throughout this paper, we assume a distance of $100\,{\rm Mpc}$ to the Coma cluster, corresponding to a distance modulus of $m-M=35.0$. All magnitudes are on the AB system and have been corrected for extinction.

\section{Data} 
\label{sec:data}

DF44 was observed with HST as part of Cycle 31 program \#17598 (PI van Dokkum) using the WFC3/UVIS channel. The observations were obtained in two epochs, in February and December 2025. The dataset consists of 3 orbits in the F200LP filter and 9 orbits in the F350LP filter. We use only the F350LP imaging: although F200LP adds $\sim 15$\,\% in signal-to-noise (S/N), combining the two filters introduced artifacts in the drizzled mosaic and led to the rejection of compact sources during cosmic ray cleaning. The F350LP data comprise 24 exposures of $\sim$1250\,s each, for a total of $\sim$30{,}000\,s. A standard dither pattern improved point spread function (PSF) sampling, filled the inter-chip gap, and facilitated rejection of cosmic rays and detector artifacts. F350LP is an extremely broad ``white-light'' long-pass filter ($\lambda$ = 3000 -- 10000 \AA), far more efficient at collecting photons from faint compact sources than the narrower filters typically used for GC studies. For an old, metal-poor point source at the GC luminosity function (GCLF) turnover ($m_V \sim 27.6$ mag), our 9 orbits reach a point-source depth equivalent to $\sim$18 orbits in F606W.

We retrieved the calibrated, charge-transfer-efficiency corrected individual exposures from the Mikulski Archive for Space Telescopes. Before combining the frames, we estimated and removed the sky background from each exposure independently, to prevent large scale gradients introduced during flat fielding from propagating into the final mosaic. For each chip, we constructed a two dimensional sky model with the \texttt{photutils} \texttt{Background2D} routine \citep{Bradley_24}, using $200\times200$ pixel boxes, a $3\times3$ median filter, and iteratively $3\sigma$-clipped medians, after masking all sources in the segmentation image and an elliptical aperture centered on DF44. The best fit map was subtracted from each exposure.

The sky subtracted exposures were combined into a single mosaic using \texttt{AstroDrizzle} \citep{Fruchter_02, Hack_12}. We adopted a pixel scale of $0.04\,{\rm arcsec\,pixel^{-1}}$, North up orientation, a $6500\times6500$ pixel grid centered on DF44, the \texttt{minmed} combination algorithm, and cosmic ray detection with S/N thresholds of 3.5 and 2.0, scale factors of 1.2 and 0.7, and inter image rejection thresholds of $5.0\sigma$ and $3.5\sigma$. After drizzling, we applied a final global sky subtraction with the same procedure to remove any remaining low level residuals, and restricted the data to the highest-S/N regions, excluding image boundaries, chip gaps, and areas affected by artifacts.

Figure~\ref{fig:DF44_HST} compares the new F350LP mosaic to the previous ACS imaging used by \citet{vanDokkum_17} and reanalysed by \citet{Saifollahi_21}. The improvement is immediately visible: the new data recover both lower surface brightness structure and a much larger population of faint compact sources. The final field-of-view used for source detection and photometry is shown in Figure~\ref{fig:radial}, and the reduced image is available on Zenodo via \dataset[doi: 10.5281/zenodo.20722270]{\doi{10.5281/zenodo.20722270}}.

\section{The globular cluster system of DF44} 
\label{sec:gcsys}

\begin{figure*}
\centering
\includegraphics[width=\textwidth]{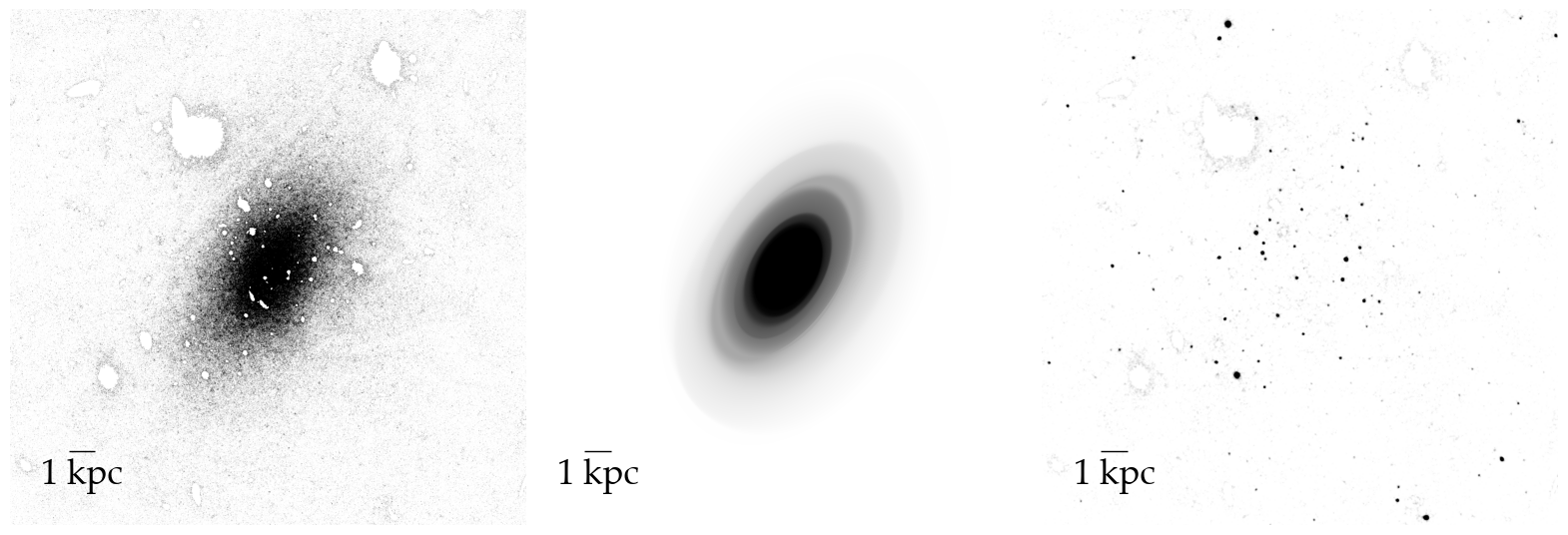}
\caption{Galaxy subtraction of DF44. \textit{Left:} Sky subtracted F350LP mosaic after masking compact sources and background galaxies. \textit{Middle:} Best fit elliptical isophotal model constructed with \texttt{photutils} \texttt{Ellipse}. 
\textit{Right:} Smoothed residual image after subtraction of the model and masking of all sources that are not well fitted by a PSF model. Numerous compact sources become clearly visible after subtraction.}
\label{fig:residual}
\end{figure*}

The controversy surrounding DF44 has largely been driven by the difficulty of measuring its GC system from shallow imaging. At 100 Mpc, the clusters are faint, their spatial extent is not known a priori, and they are projected against both the diffuse stellar body and a non-negligible background of compact sources, requiring large corrections (typically a factor of $\sim 4$) to the directly-detected counts. The new F350LP data reach well below the GCLF turnover, reducing the completeness corrections and allowing an empirical assessment of the radial distribution. In this section we construct the GC candidate catalogue and measure the spatial distribution, luminosity function, and total number of GCs.

\subsection{Galaxy fitting}

To detect faint compact sources projected against the diffuse stellar body of DF44, we modeled and subtracted the galaxy light. The galaxy was fitted with elliptical isophotes using the \texttt{Ellipse} routine within \texttt{photutils} \citep{Jedrzejewski_87, Bradley_24}. Prior to fitting, we masked all pixels flagged in a \texttt{SExtractor} \citep{Bertin_96} segmentation map, removing sources that would otherwise bias the isophote fits, and subtracted the best fitting model from the image. The resulting smoothed residual image is shown in Figure~\ref{fig:residual}.

\subsection{Source detection and GC selection}

Compact sources were detected on the galaxy subtracted residual image and measured on the original sky-subtracted F350LP science image using \texttt{SExtractor} in dual image mode \citep{Bertin_96}. 
We ran \texttt{SExtractor} with PSF fitting photometry enabled, which provides higher S/N ratio measurements for compact sources than fixed aperture methods.

We modeled the PSF of the F350LP mosaic using \texttt{PSFEx} \citep{Bertin_11}, building an empirical PSF from a sample of bright, isolated, unsaturated point sources distributed across the field. 
We initially ran \texttt{SExtractor} in single image mode on the galaxy-subtracted residual image, but found that PSF magnitudes were sometimes impacted by non-Gaussian residuals from the galaxy fit.
The dual image mode avoids biases from non-Gaussian galaxy-fit residuals in the PSF magnitudes while maintaining sensitivity to faint sources within the body of the galaxy; aperture magnitudes measured in the two images are nearly identical, confirming that the local background is measured correctly.

We selected GC candidates as compact sources with a finite PSF-fit magnitude and
$2.0 ($0\farcs08$) < \mathrm{FWHM} < 4.1 ($0\farcs16$)$\,pixels. Requiring a finite PSF magnitude removes objects poorly described by the empirical PSF (extended background galaxies, blends, and detections affected by residual structure), while the FWHM cut isolates sources that are at most marginally resolved at the distance of Coma.

This selection yields 98 GC candidates and includes all visually apparent compact objects, although a few genuine clusters may be lost to strong galaxy subtraction residuals or blending with background galaxies. The photometry was converted from F350LP to Johnson $V$ using the Flexible Stellar Population Synthesis code \citep{Conroy_09, Conroy_10}. We generated a grid of 13\,Gyr, dust-free, Kroupa initial mass function stellar populations spanning $-2.0\leq $[M/H]$ \leq-0.5$, weighted by a Gaussian centered at [M/H]$=-1.5$ with $\sigma=0.3$\,dex, representative of old, metal-poor GCs. The resulting transformation, $m_V = m_\mathrm{F350LP} - 0.028$\,mag, varies by less than $0.01$\,mag across the grid, negligible compared to the photometric uncertainties, and was applied to all F350LP magnitudes. We verified this calibration empirically using archival WFC3/UVIS F606W imaging of DF44: for common point sources, the F350LP photometry agrees with F606W to better than $0.1$\,mag. As an independent check, we repeated the full analysis using aperture rather than PSF photometry and recovered consistent values of $R_{\rm GC}/R_{\rm e}$, $N_{\rm GC}$, the GCLF turnover magnitude, and $\sigma$.

Of the 28 candidates catalogued for DF44 by \citet{Saifollahi_21}, 21 are recovered with consistent photometry in our study; the remaining 7 fail our FWHM cut, indicating extended or blended sources, most likely background galaxies. Due to the same reasons, of the 21 candidates in \cite{Saifollahi_22}, we find 17. Our GC selection and photometric measurements were also compared with \cite{vanDokkum_16}. Although individual source magnitudes were not tabulated in that study, their published GCLF shows that the four brightest objects occupy the same magnitude bins as in our catalog. Notably, the two brightest sources in our catalog are detected in all three studies. The brightest, with a magnitude of $V = 23.84 \pm 0.01$, has been flagged as a likely Milky Way star by \cite{vanDokkum_19_DF44} using Keck/KCWI spectroscopy, while the second brightest lies at the boundary of the ultra-compact-dwarf (UCD) regime ($M_V \approx -11$).

\subsection{Completeness and artificial star tests}
\label{sec:artificial}

We characterised the photometric completeness through artificial star tests. We injected 10000 scaled \texttt{PSFEx}-model point sources with $22\leq m_V \leq30$ mag at random valid positions, added simultaneously to the detection and measurement images, reran \texttt{SExtractor} with an identical configuration, and matched recovered sources to injected positions within 3 pixels. 

The overall recovery fraction across the full magnitude range is $77.4\%$. We fitted the completeness as a function of magnitude following \citet{Janssens_22}, and find a 50\,\% completeness magnitude of $m_{50}~=~29.32$\,mag, corresponding to $M_V~\sim~-5.7$ mag at 100\,Mpc. The completeness is $>90\%$ for $m_V~<~28.9$\,mag, or $M_V~<~-6.1$ mag, and drops steeply between $m_V~=~29.1$ and $m~=~29.9$\,mag. Critically, the canonical GCLF turnover at $M_V~=~-7.4$ mag \citep[e.g.,][]{Harris_01} lies well above the 50\% completeness threshold, so the GC population can be measured without relying on large incompleteness corrections.

\subsection{Radial surface density profile} 
\label{sec:radial}

\begin{figure*}
\centering
\includegraphics[width=0.93\textwidth,trim=0 0 0 1.0cm,clip]{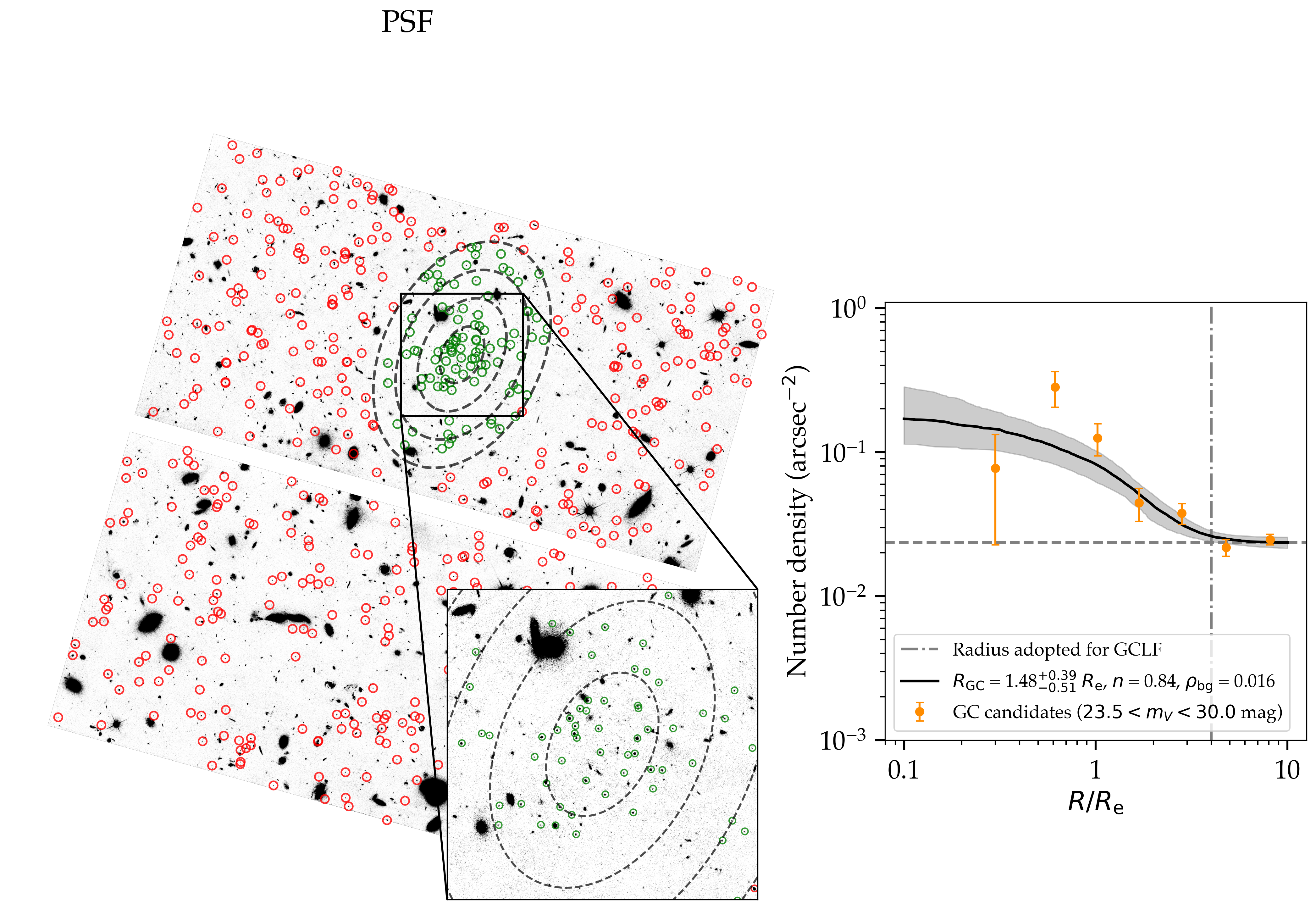}
\caption{Spatial distribution and radial surface density profile of GC candidates around DF44. \textit{Left:} Residual F350LP mosaic showing the compact sources selected as GC candidates within $4R_{\rm e}$ in green, and sources satisfying the same selection criteria outside $4R_{\rm e}$ in red. Dashed ellipses indicate multiples of the galaxy effective radius. The inset highlights the central region of the galaxy. \textit{Right:} Radial surface density profile measured in elliptical annuli aligned with the stellar body of DF44. Orange circles show the raw number density measurements, and the solid black curve shows the best fit Sérsic model obtained with \texttt{emcee}. Radii are measured along the semi-major axis, in units of the stellar $R_{\rm e}$. The dashed vertical line shows the adopted radius inside of which we selected the GCs that will contribute to the GCLF. The fitted profile implies an extended GC system with $R_{\rm GC}=1.48^{+0.39}_{-0.51}R_{\rm e}$ and Sérsic index $n_{\rm GC}=0.84^{+0.37}_{-0.24}$. The dashed horizontal line shows the best-fitting background density of $\rho_{\rm bg} = 0.016^{+0.012}_{-0.006}$\,arcsec$^{-2}$.}
\label{fig:radial}
\end{figure*}

The surface density of GCs was measured in elliptical annuli aligned with the photometric major axis of DF44, adopting PA$=-26.4^\circ$ and $b/a=0.66$ \citep[][reproduced by our \texttt{Ellipse} fitting]{vanDokkum_17}. The annuli were logarithmically spaced, with areas corrected for chip gaps and field edges. The GC spatial distribution and resulting radial profile are shown in Figure~\ref{fig:radial}.

We fitted the profile with a Sérsic function using the \texttt{emcee} Markov Chain Monte Carlo sampler \citep{Foreman-Mackey_13}, treating the Sérsic index, half-number radius ($R_{\rm GC}$), normalization, and background density as free parameters. The best fit yields a S\'ersic index $n_{\rm GC}~=~ 0.84^{+0.37}_{-0.24}$ and a half-number radius $R_{\rm GC}~=~ 1.48^{+0.39}_{-0.51}\,R_{\rm e}$ ($11.6^{+3.1}_{-4.4}$ arcsec, or $5.6^{+1.5}_{-2.1}$ kpc). The fitted background density is $\rho_{\rm bg}~=~0.016^{+0.012}_{-0.006}\,{\rm arcsec^{-2}}$. The GC system is therefore more extended than the stellar body, with a substantial fraction of the population outside $1R_{\rm e}$, making the adopted radial aperture a potentially dominant systematic in shallow data.

\subsection{GCLF and GC total number}

\begin{figure*}
\centering
\includegraphics[width=\textwidth]{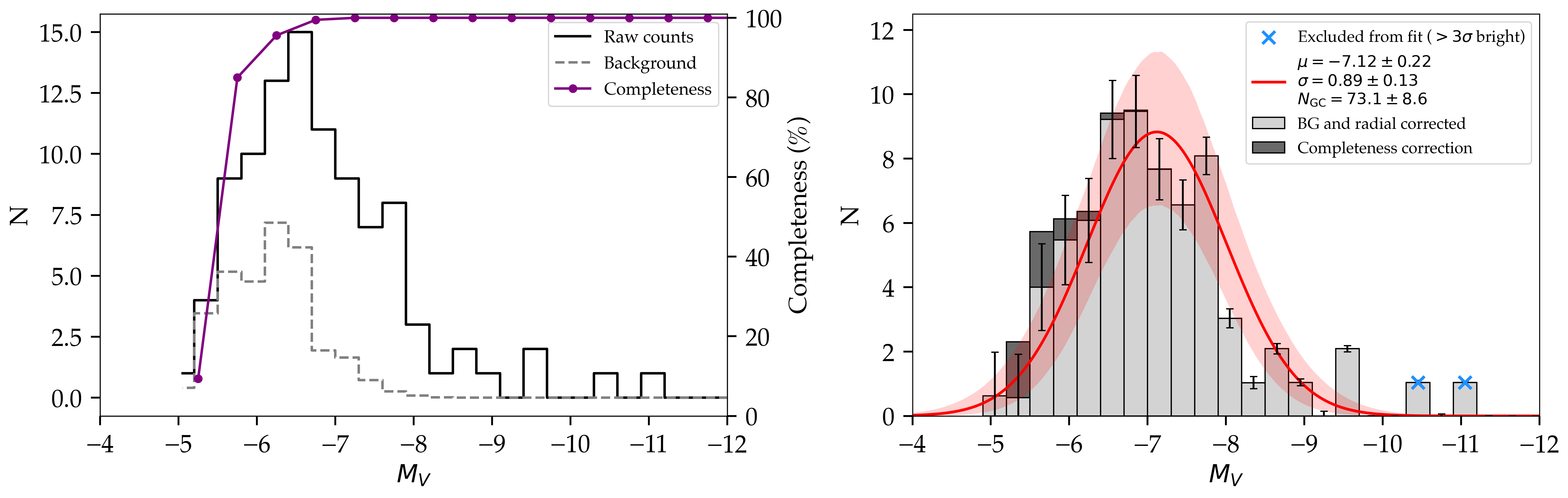}
\caption{GCLF of DF44. \textit{Left:} Raw GC candidate counts, estimated background counts, and completeness curve derived from artificial star tests. \textit{Right:} Background-subtracted and completeness-corrected GCLF. Light gray bars show the background-corrected counts, while the dark gray component indicates the additional correction applied for incompleteness. The red curve shows the Gaussian GCLF obtained from the unbinned fit, with a turnover magnitude of $M_{V,\mathrm{TO}}=-7.12\pm0.22$ mag and a dispersion of $\sigma=0.89\pm0.13$ mag. The two brightest bins, which contain likely outliers, are excluded from the fit. Summing the corrected bins gives a total population of $N_{\rm GC}=73.1\pm8.6$.}
\label{fig:GCLF}
\end{figure*}

The GCLF was constructed from the compact sources within $4R_{\rm e}$ of DF44, binned in $0.3$ magnitude intervals (Fig.~\ref{fig:GCLF}). The background was estimated per magnitude bin from the same catalogue, using all sources that satisfy the GC selection criteria but lie beyond $5R_{\rm e}$ of DF44. The counts in each bin were scaled to the GC counting region by the ratio of the two areas, both measured over the usable footprint of the mosaic. To test whether this background is representative, we measured the density of selected sources in independent, non-overlapping apertures distributed across the field. The field-to-field scatter is consistent with Poisson statistics for the brighter sources and exceeds it by ${\sim}30\%$ at the faintest magnitudes, and we propagate this scatter into the uncertainties on the background-subtracted counts. The resulting background density agrees with the value fitted independently as a free parameter of the radial profile ($\rho_{\rm bg}=0.016^{+0.012}_{-0.006}\,{\rm arcsec}^{-2}$; Section~\ref{sec:radial}). Each bin was then corrected for the expected background contamination, for the expected fraction of GCs outside of $4R_{\rm e}$, and for the completeness at that magnitude.

Summing all corrected bins yields $N_{\rm GC}=73.1\pm8.6$. This confirms that DF44 is among the most GC-rich UDGs known and is consistent with the original estimate from shallower HST imaging by \citet{vanDokkum_17}. Without any correction other than background subtraction, the number of directly measured GCs is $66$. Our corrections add only $\sim7$ GCs, and the galaxy is GC-rich whether or not they are applied.

As a consistency check, integrating the number density profile (Fig.~\ref{fig:radial}) gives $N_{\rm GC}=56.3^{+9.3}_{-9.1}$, or $N_{\rm GC}\sim73$ after accounting for the $77.4\%$ photometric completeness, consistent with the GCLF result. Restricting the measurement to $R<0.8R_{\rm e}$ gives $N_{\rm GC}=12.6^{+3.3}_{-3.1}$ ($\sim16.5$ after completeness). Therefore, even assuming a GC distribution more compact than the stellar light, the total is $\gtrsim32$.

We fitted a Gaussian to the unbinned data to recover the GCLF turnover and dispersion \citep{Jordan_07,Villegas_10,Mueller_21,Saifollahi_25b}. We modelled the observed magnitude distribution as the sum of a Gaussian GCLF and an empirical background component, fitting both components simultaneously. The fit uncertainties account for Poisson noise, individual photometric errors, magnitude-dependent incompleteness, and the uncertainty in the background inferred from the measured field-to-field scatter. The final fit gives a turnover magnitude of $M_{V,\mathrm{TO}}=-7.12\pm0.22$ mag and a dispersion of $\sigma=0.89\pm0.13$ mag, with the two brightest sources excluded from the likelihood fit (as shown in Fig. \ref{fig:GCLF}). The turnover is slightly fainter than the values adopted by \citet{vanDokkum_16} and \citet{Saifollahi_22}, although the measurements remain consistent within the uncertainties. Notably, their fits included the two brightest sources excluded from our fit, one of which is a known Milky Way star \citep{vanDokkum_19_DF44}, while the other is likely a UCD. Removing these two sources from their fits would naturally shift their inferred turnovers toward fainter magnitudes. In fact, a fainter turnover is not unexpected for a dwarf galaxy. Studies of dwarf galaxies in Fornax \citep{Villegas_10,Jordan_15,Saifollahi_25}, Virgo \citep{Villegas_10}, Perseus \citep{Saifollahi_25b}, and other nearby environments \citep{Miller_Lotz_07,Georgiev_09} show that, on average, the GCLF turnover becomes systematically fainter toward lower-luminosity hosts. This likely reflects a lower characteristic GC mass together with differences in the dynamical evolution, ages, and metallicities of their cluster populations \citep[see][for a thorough discussion of dwarf galaxy GCLFs]{Gannon_26}.

\section{Discussion}
\label{sec:discussion}

\subsection{The GC population of DF44}

The new imaging resolves the main observational question surrounding the GC system of DF44. We find $N_{\rm GC}=73.1\pm8.6$ and $S_N=42.1\pm4.9$, placing DF44 among the most GC-rich galaxies known for its luminosity \citep[see e.g., ][]{Pfeffer_24}. Importantly, this result does not depend strongly on corrections to the faint end of the GCLF: after background subtraction alone, we directly measure 66 GCs. The new data therefore favor the high end of the previously published estimates and establish that the unusually rich GC system of DF44 is a real property of the galaxy.

Part of the remaining difference with \citet{Saifollahi_21,Saifollahi_22} may be connected to the inferred spatial extent of the GC system. They measured $R_{\rm GC}/R_{\rm e}~\sim~0.8$, whereas we find $R_{\rm GC}/R_{\rm e}=1.48^{+0.39}_{-0.51}$. The importance of $R_{\rm GC}$ for the inferred total population was already discussed in detail by \citet{Saifollahi_21,Saifollahi_22}, who showed that the radial distribution is one of the main uncertainties in measuring GC populations at the distance of Coma. In their data, the GC surface density around DF44 approaches the background at approximately $2R_{\rm e}$, making an extended low-density population difficult to constrain.

The radial extent alone, however, does not explain the full difference between the published GC numbers. Even within $R<0.8R_{\rm e}$, approximately the half-number radius inferred by \citet{Saifollahi_21,Saifollahi_22}, we find ${\sim}16$ GCs after background and completeness correction. If this radius enclosed half of the population, the corresponding total would already be ${\sim}32$ GCs. Differences in source selection, background estimation, and the magnitude range used to constrain the population can therefore also contribute. This is particularly important when the GC density becomes comparable to the background, as small changes in either quantity can propagate into the inferred radial profile and total number. The deeper data reduce these dependencies by measuring the GCLF beyond its turnover and tracing the GC distribution farther into the regime where it approaches the background.

\subsection{What does the GC system tell us about the formation of DF44?}

If DF44 follows the empirical $N_{\rm GC}$--halo mass relation \citep{Harris_17,Burkert_Forbes_20}, its GC population implies $\log(M_{\rm vir}/M_\odot)~=~11.5~\pm~0.3$. This is consistent within the uncertainties with the massive, cored-halo solution allowed by the spatially resolved stellar kinematics of \citet{vanDokkum_19_DF44}, who found $\log(M_{200}/M_\odot)=11.2\pm0.6$, and lies $\sim$1\,dex above their cuspy NFW fit ($\log M_{200} = 10.6^{+0.4}_{-0.3}$). We therefore take the agreement as independent support for a relatively massive cored dark matter halo around DF44. For $M_\star=3\times10^8\,M_\odot$, the GC-inferred halo mass leads to $M_\star/M_{\rm vir}\sim10^{-3}$, placing DF44 well below the stellar-to-halo mass ratio of normal dwarf galaxies, with a dark matter fraction $>99.9\%$.

The GC mass fraction provides complementary information. Integrating the completeness-corrected GCLF and adopting $M/L_V=2$, we find the mass in GCs to be $M_{\rm GC}=(1.2\pm0.1)\times10^7\,M_\odot$, or $M_{\rm GC}/M_\star\sim3.9\%$ (assuming a stellar mass of $M_{\star} = 3 \times 10^8 M_{\odot}$, and excluding from the calculation the two brightest sources in our catalogue, which are a MW star and a likely UCD). This is substantially higher than the $\lesssim1\%$ typical of dwarf galaxies and lies above the proposed ${\sim}2.5\%$ division between formation pathways discussed by \cite{Forbes_25}, and also argued as a sign of early formation in \citealt{Saifollahi_22}. One of the clearest examples of a failed galaxy, NGC~5846-UDG1, reaches ${\sim}13\%$ \citep{Danieli_22}, but this is not unique to failed galaxies, an example being the star-forming merging dwarf UGC~9050-Dw1 reaching ${\sim}21\%$ \citep{Fielder_23}.

The interpretation of DF44 therefore does not rest on any one of these quantities. It is their combination that is unusual: $S_N>40$, $M_{\rm GC}/M_\star\sim3.9\%$, an old and metal-poor stellar population, a short formation timescale, little or no rotation, and a high inferred halo mass. Together these properties strongly favor the failed-galaxy interpretation, in which DF44 formed efficiently at early times but built relatively little field-star mass afterward. This is also consistent with the spatially resolved stellar populations. \citet{Villaume_22} found elevated, GC-like $[\mathrm{Mg/Fe}]$ in the central ${\sim}1.5$ kpc and lower values at larger radii, together with flat age and flat-to-positive metallicity gradients. Other stellar-population studies similarly have found that most of the stellar mass formed early and that star formation ended rapidly \citep{Webb_22,FerreMateu_23}. The formation of DF44 was therefore likely more structured than a single instantaneous burst, but still early and relatively rapid.

The question is then what formation scenario can produce this combination of properties. Several mechanisms can make dwarf galaxies diffuse, including high halo spin, bursty stellar feedback, environmental processing, and mergers \citep{Amorisco_16,diCintio_17,Chan_18,Carleton_19,Sales_20,Tremmel_20}. These models can reproduce the large size of DF44, but the additional constraints from its rich GC system, early quenching, and high inferred halo mass make the problem more restrictive.

Current simulations reproduce parts of this picture, although not yet the full combination of properties seen in DF44. TNG50 produces UDGs with up to ${\sim}30$ GCs, but does not contain systems as GC rich as DF44 \citep{Doppel_23}. In E-MOSAICS, GC-rich dwarf galaxies form preferentially at early times and high gas pressures, and their enclosed masses overlap those measured for observed GC-rich UDGs, including DF44 \citep{Pfeffer_24}. The simulations nevertheless predict that GC-rich dwarfs should live in undermassive halos, the opposite of what is found for DF44. A search in the Magneticum simulations similarly identified possible high-redshift progenitors of failed galaxies, but no present-day systems matching their full set of properties \citep{Gannon_25}. None of these simulations contains a Coma-mass cluster, so the absence of a direct DF44 analogue may also reflect the rarity of these systems and the limited simulated volumes. Recent idealized simulations show that gas-rich dwarf mergers can produce dispersion-supported UDGs with 20--40 massive clusters \citep{Tapia_26}. Their remnants, however, retain the cuspy dark matter profiles of their progenitors and host GC systems that are more centrally concentrated than the field stars, the opposite of the cored halo and spatially extended GC system we find for DF44. 

Models that connect massive-cluster formation to the high gas pressures of early star formation provide a possible way forward. In the model of \citet{Trujillo-Gomez_22}, high gas pressures at early times lead to efficient formation of massive bound clusters, while the associated clustered feedback can expand both the stellar and dark matter distributions. This provides a natural connection between a rich GC population and an intense early phase of star formation, and was also invoked by \citet{Danieli_22} to explain NGC~5846-UDG1. \citet{Trujillo-Gomez_22} showed that this process could account for several properties of DF44, including its large size, old stellar population, and relatively flat stellar-population gradients, but does not account for the high velocity dispersion of DF44, as the model generally leads to low velocity dispersion dwarfs, like the dark matter deficient UDG counterparts.

These models therefore can reproduce some of the observed properties of DF44, but they do not yet explain the full combination of its rich GC system, high inferred halo mass, and low stellar mass. With the GC population now much better constrained, the main question is no longer how many clusters DF44 contains, but what early conditions allowed it to form so many GCs while forming comparatively few field stars.

\section{Conclusions} 
\label{sec:conclusions}

We have presented new deep \textit{HST}/WFC3 imaging of DF44 in the broad F350LP filter, reaching more than one magnitude below the expected GCLF turnover. These data provide the deepest census to date of the compact sources around DF44 and allow us to revisit whether it is truly an extreme GC-rich failed galaxy.

We find that the GC system of DF44 is spatially extended, with a half-number radius of $R_{\rm GC}=1.48^{+0.39}_{-0.51}R_{\rm e}$ ($5.6^{+1.5}_{-2.1}$ kpc). The unbinned GCLF fit gives a turnover magnitude of $M_{V,\mathrm{TO}}=-7.12\pm0.22$ mag and a dispersion of $\sigma=0.89\pm0.13$ mag. This turnover is slightly fainter than the value typically adopted for massive galaxies, but consistent with the range observed in dwarf galaxies. The background-subtracted and completeness-corrected GCLF gives $N_{\rm GC}=73.1\pm8.6$ and a specific frequency of $S_N=42.1\pm4.9$, confirming DF44 as one of the most GC-rich UDGs known, with a GC system unusually massive for its stellar mass.

The $N_{\rm GC}$--halo mass relation implies $\log(M_{\rm vir}/M_\odot)=11.5\pm0.3$, consistent with the cored-halo fit of \citet{vanDokkum_19_DF44} based on spatially resolved stellar kinematics and independently supporting a high halo mass for DF44. This places DF44 among the UDGs that appear overmassive for their stellar content, with a dark matter fraction above $99.9\%$. Its high GC system mass fraction of $M_{\rm GC}/M_\ast=3.9\%$ further reinforces GC formation efficiency as one of the most robust tracers of the failed-galaxy pathway.

Taken together, the new data confirm that DF44 is one of the clearest GC-rich failed-galaxy candidates known. Its rich and extended GC system, high $S_N$, high $M_{\rm GC}/M_\star$, and high GC-inferred halo mass show that it is not simply a normal dwarf made diffuse. The remaining question is not whether DF44 is GC rich, but what physical conditions allowed it to form such a massive GC system while building so little diffuse stellar mass.

\begin{acknowledgments}
We are thankful to the referee, as well as Duncan Forbes and Teymoor Saifollahi, for comments and suggestions which greatly improved this manuscript.

This work is based on observations made with the NASA/ESA Hubble Space Telescope, obtained at the Space Telescope Science Institute, which is operated by the Association of Universities for Research in Astronomy, Inc., under NASA contract NAS5-26555. 
AJR was supported by National Science Foundation grant AST-2308390.

These observations are associated with program \#17598, PI van Dokkum. Raw data can be accessed from the Mikulski Archive for Space Telescopes (MAST) at the Space Telescope Science Institute via \dataset[doi: 10.17909/jj23-vj59]{\doi{10.17909/jj23-vj59}}. The final reduced image of DF44 used in this work and a table with coordinates and magnitudes of all GC candidates are available on Zenodo via DOI \dataset[10.5281/zenodo.20722270]{\doi{10.5281/zenodo.20722270}}.
\end{acknowledgments}

\bibliography{DF44}{}
\bibliographystyle{aasjournalv7}

\end{document}